\documentclass{article}

\usepackage{microtype}
\usepackage{graphicx}
\usepackage{subfigure}
\usepackage{booktabs} % for professional tables

\usepackage{hyperref}

\usepackage[accepted]{icml2025}

\usepackage{amsmath}
\usepackage{amssymb}
\usepackage{mathtools}
\usepackage{amsthm}
\usepackage{algorithm}
\usepackage{algorithmic}
\usepackage{graphicx}

\usepackage[capitalize,noabbrev]{cleveref}

\theoremstyle{plain}

\theoremstyle{definition}

\theoremstyle{remark}

\usepackage[textsize=tiny]{todonotes}
\newcommand{\etal}{\textit{et al.}}
\newcommand{\eg}{\textit{e.g.}}

\begin{document}
	
	\twocolumn[
	\icmltitle{Vanishing Watermarks: Diffusion-Based Image Editing Undermines Robust Invisible Watermarking}
	
		\begin{icmlauthorlist}
		\icmlauthor{Fan Guo}{}
		\icmlauthor{Jiyu Kang}{}
		\icmlauthor{Qi Ming}{}
		\icmlauthor{Emily Davis}{}
		\icmlauthor{Finn Carter}{}
	\end{icmlauthorlist}
	
	\begin{icmlauthorlist}
		{Xidian University}
	\end{icmlauthorlist} 
	
	%
	% \icmlcorrespondingauthor{Firstname1 Lastname1}{first1.last1@xxx.edu}
	% \icmlcorrespondingauthor{Firstname2 Lastname2}{first2.last2@www.uk}
	
	% You may provide any keywords that you
	% find helpful for describing your paper; these are used to populate
	% the "keywords" metadata in the PDF but will not be shown in the document
	\icmlkeywords{Machine Learning, ICML}
	
	\vskip 0.3in
	]
	
	% this must go after the closing bracket ] following \twocolumn[ ...
	
	% This command actually creates the footnote in the first column
	% listing the affiliations and the copyright notice.
	% The command takes one argument, which is text to display at the start of the footnote.
	% The \icmlEqualContribution command is standard text for equal contribution.
	% Remove it (just {}) if you do not need this facility.
	
	% \printAffiliationsAndNotice{}  % leave blank if no need to mention equal contribution
	% \printAffiliationsAndNotice{\icmlEqualContribution} % otherwise use the standard text.
\begin{abstract}
	Robust invisible watermarking schemes aim to embed hidden information into images such that the watermark survives common manipulations. However, powerful diffusion-based image generation and editing techniques now pose a new threat to these watermarks. In this paper, we present a comprehensive theoretical and empirical analysis demonstrating that diffusion models can effectively erase robust watermarks even when those watermarks were designed to withstand conventional distortions. We show that a diffusion-driven {\it image regeneration} process, which leverages generative models to recreate an image, can remove embedded watermarks while preserving the image’s perceptual content
	. Furthermore, we introduce a guided diffusion-based attack that explicitly targets the embedded watermark signal during generation, significantly degrading watermark detectability. Theoretically, we prove that as an image undergoes sufficient diffusion transformations, the mutual information between the watermarked image and the hidden payload approaches zero, leading to inevitable decoding failure
	. Experimentally, we evaluate multiple state-of-the-art watermarking methods (including deep learning-based schemes like StegaStamp, TrustMark, and VINE) and demonstrate that diffusion edits yield near-zero watermark recovery rates after attack, while maintaining high visual fidelity of the regenerated images
	. Our findings reveal a fundamental vulnerability in current robust watermarking techniques against generative model-based edits, underscoring the need for new strategies to ensure watermark resilience in the era of powerful diffusion models.
\end{abstract}

%%%%%%%%% INTRODUCTION
\section{Introduction}
Digital watermarking enables hidden messages to be embedded in images for applications such as copyright protection, provenance tracking, and content authenticity verification. In particular, {\em robust invisible watermarking} focuses on encoding information imperceptibly such that it remains recoverable even after various image manipulations (e.g., cropping, compression, or noise). Classical approaches based on signal processing and information coding (e.g., spread-spectrum or frequency-domain methods) established early techniques for imperceptible yet robust marks. Modern learned watermarking schemes, such as deep neural network-based methods, have significantly advanced robustness against common distortions by jointly optimizing encoder and decoder networks. For example, Tancik \etal introduced {\sc StegaStamp}, a learned steganographic algorithm that hides a bitstring in images and can reliably recover it even after printing, scanning, or camera-capture processes \cite{Tancik2020}. These systems achieve impressive invariance to many conventional perturbations.

However, the emergence of powerful diffusion models for image generation and editing has created a new category of {\em generative} transformations that robust watermarks were not explicitly designed to survive
. Diffusion-based image editors can produce photorealistic modifications of an image or even entirely regenerate an image’s content, potentially obliterating the low-level details that encode a watermark while preserving high-level semantics. This raises an urgent question: \textit{Are robust invisible watermarks fundamentally vulnerable to removal by diffusion-based image editing?} Recent anecdotal evidence and preliminary studies suggest that generative models (e.g., text-to-image diffusion) can indeed “wash out” watermarks by re-synthesizing the image content anew, akin to making the watermark {\em vanish}.

In this work, we provide a thorough investigation of how diffusion-based image editing processes can unintentionally (or intentionally) compromise robust watermarks. We consider a threat model where an adversary has access to a diffusion model (such as Stable Diffusion) and uses it to manipulate a watermarked image, either through image-to-image generation, inpainting, or other editing techniques, with the goal of preserving the image’s visual content but removing the hidden watermark payload. We examine both unguided diffusion regeneration (where the image is simply regenerated with similar appearance) and {\em guided attacks} where the diffusion process is steered to specifically disrupt the watermark signal. Both scenarios reflect realistic use cases: a benign user might unknowingly remove a watermark by applying an AI image filter, whereas a malicious actor might deliberately attempt to scrub a watermark for unauthorized use of content.

Our contributions are summarized as follows. \textbf{(1)} We demonstrate empirically that leading robust watermarking methods—\eg, {\sc StegaStamp} \cite{Tancik2020}, {\sc TrustMark} \cite{Bui2023}, and {\sc VINE} \cite{lu2024robust}—exhibit a dramatic drop in decoding accuracy after diffusion-based image editing. In our experiments, diffusion edits often reduce watermark recovery to near-zero even though these schemes withstand standard perturbations (Section \ref{sec:results}). \textbf{(2)} We propose a novel {\em diffusion watermark removal} algorithm that integrates a watermark decoder’s feedback into the diffusion model’s denoising loop to explicitly target and erase the hidden message (Section \ref{sec:method}). \textbf{(3)} We provide a theoretical analysis explaining why diffusion transformations are particularly effective at breaking watermarks. We formally show that as the diffusion process progresses and approaches an image’s prior distribution, the mutual information between the image and any embedded watermark bitstream tends to zero (Section \ref{sec:theory}). This analysis builds on information-theoretic arguments to confirm that sufficiently powerful generative edits inevitably destroy embedded signals. \textbf{(4)} We discuss the broader implications of our findings for the design of future watermarking schemes (Section \ref{sec:discussion}). In particular, we highlight ethical concerns: content authenticity initiatives relying on watermarks (such as those by the Content Authenticity Initiative) may be undermined by generative AI tools, and robust watermarking must evolve to address attacks by learned generative models.

Overall, our work reveals a pressing vulnerability at the intersection of two rapidly developing areas: diffusion-based image generation and robust watermarking. We hope this study will spur new research into watermark techniques that can survive AI-based content manipulation, as well as mitigations (like watermark detectors or model fingerprinting) to maintain trust in digital imagery in the age of generative AI.

\section{Related Work}
\label{sec:related}

\paragraph{Robust Invisible Watermarking.}
Digital watermarking~\cite{ren2025all} has long been studied as a means to embed information into media in a way that is invisible yet resilient to distortions. Early techniques (e.g., Cox \etal’s spread-spectrum method in the DCT domain) demonstrated that a few bits could be hidden and later extracted even after attacks like compression or noise. With the rise of deep learning, recent approaches have greatly increased capacity and robustness. \textbf{Deep watermarking networks} train an encoder-decoder pair: an encoder network hides a message in an image, and a decoder network recovers the message from possibly perturbed images. Tancik \etal’s {\sc StegaStamp} \cite{Tancik2020} pioneered this approach, achieving 56-bit payloads that survive camera captures with small errors. Secure encoding methods like adversarial training and error-correcting codes have further improved robustness. {\sc RivaGAN} introduced attention-based embedding for robust video watermarking, and its image adaptation also showed resilience to many filters. More recently, {\sc TrustMark} by Bui \etal \cite{Bui2023} proposed a GAN-based watermarking architecture with a novel spatio-spectral loss to balance imperceptibility and robustness. TrustMark’s decoder is designed to handle arbitrary resolution inputs and a wide range of perturbations, and it even includes a proposed removal network for cases where re-watermarking is needed (i.e., to intentionally remove a watermark for replacement). Another notable scheme is {\sc VINE} (Watermarking using generative priors) by Lu \etal \cite{lu2024robust}, which specifically addresses robustness against {\em generative} image edits. VINE introduces a comprehensive benchmark (W-Bench) evaluating watermark survival through various image editing scenarios (global edits, local edits, image-to-video, etc.) and finds that most existing methods fail under these conditions. To improve robustness, VINE leverages a large diffusion model (SDXL) as part of the watermark encoder to better hide signals in imperceptible ways, and trains with simulated diffusion-based attacks (using blur as a surrogate for generative distortions) to make the watermark harder to remove
. While VINE outperforms prior schemes under many editing operations, our results (Section \ref{sec:results}) indicate that even VINE’s advanced design can be overcome by powerful diffusion regeneration attacks, suggesting an arms race between watermark embedding and generative removal.

\paragraph{Diffusion-Based Image Editing.}
Diffusion models, such as DDPMs and the latent diffusion backbone of Stable Diffusion \cite{Rombach2022}, have revolutionized image generation and enabled rich image editing capabilities. By modeling the distribution of natural images, diffusion models can remove or alter image content via a generative denoising process guided by user inputs (text prompts, reference images, etc.). A multitude of {\em text-driven} editing techniques have been proposed: e.g., prompt-to-prompt editing \cite{Hertz2022} modulates cross-attention to achieve specified changes, and null-text inversion \cite{Mokady2023} inverts an image to a noise latent that can be used to re-generate the image with modifications. Inpainting and outpainting with diffusion models allow content removal or addition in user-specified regions simply by conditioning on masked images. More advanced control mechanisms include scribbles, segmentation maps, or keypoint guidance (via ControlNet and related techniques) to achieve fine-grained edits. Of particular relevance are works that preserve overall image structure while altering specific aspects, since these can potentially maintain an image’s appearance (satisfying a human user) while unintentionally stripping away low-level details like watermarks. For instance, \textbf{cross-domain image composition} can insert an object from one image into another’s context. Lu \etal’s {\sc TF-ICON} \cite{Lu2023} is a training-free diffusion-based composition framework that uses an “exceptional” empty prompt to invert real images and then seamlessly blend objects, achieving superior quality without fine-tuning. Similarly, Gao \etal’s {\sc SHINE} \cite{Lu2025a} (Seamless High-fidelity Insertion with Neutralized Errors) leverages a diffusion transformer (FLUX) with manifold-steered losses and blending to insert objects with physically plausible consistency (handling shadows, reflections, etc.). Such methods aim to produce a composed image that looks natural and unedited, which inherently means any foreign signals (like watermarks) that do not semantically belong in the scene are likely to be eliminated during generation. Another line of work is \textbf{drag-based interactive editing}, where a user drags points to reposition or deform objects in the image. Zhou \etal’s {\sc DragFlow} \cite{Zhou2025} exemplifies state-of-the-art drag-based editing using diffusion transformers: it introduces region-based motion supervision and adapter-guided inversion to move objects in an image while preserving fidelity and background integrity. DragFlow surpasses prior point-based methods in producing realistic results without distortion. In doing so, it effectively re-synthesizes the dragged object in the new location from the model’s prior, again suggesting that any hidden watermark tied to the original pixel arrangement could be lost. Overall, diffusion-based editors like TF-ICON, SHINE, DragFlow, and many others \cite{Hertz2022,Mokady2023} provide unprecedented post-processing capabilities. Our study sheds light on how these beneficial tools inadvertently conflict with watermark preservation.

\paragraph{Concept Erasure in Diffusion Models.} 
Interestingly, the ability of diffusion models to remove specific information is actively being researched in the context of \emph{concept erasure}. Concept erasure techniques aim to fine-tune or guide diffusion models so that they cannot generate certain undesirable content (such as copyrighted characters, private individuals, or harmful imagery). This is directly related to our topic: a watermark can be viewed as a ``concept'' (albeit an invisible one) embedded in an image, and diffusion models can either accidentally or deliberately erase that concept during generation. Recent methods like {\sc MACE} (Mass Concept Erasure) by Lu \etal \cite{Lu2024}, {\sc ANT} (Auto-steering denoising Trajectories) by Li \etal \cite{li2025ant}, and {\sc EraseAnything} by Gao \etal \cite{Gao2025} provide ways to remove or ``forget’’ target concepts in diffusion models. MACE fine-tunes a model to erase up to 100 concepts at once while balancing generality and specificity (ensuring synonyms are also erased while unrelated outputs remain intact). ANT introduces a trajectory-aware objective that only alters the denoising trajectory in the mid-to-late stages to sidestep unwanted concepts without harming early structure, achieving state-of-the-art performance in single- and multi-concept suppression \cite{Li2025}. EraseAnything tackles concept removal in the latest diffusion architectures (like Flow Matching-based transformers) using a bi-level optimization with LoRA fine-tuning and attention map regularization, specifically addressing cases like removing a visual element (e.g., nudity) from generation \cite{Gao2025,gao2025revoking}. The relevance to watermarking is clear: these works demonstrate that diffusion models can be strategically controlled or fine-tuned to remove designated information from outputs. In our case, the ``unwanted concept’’ is the embedded watermark signal. While concept erasure methods are typically used to enforce safety or copyright constraints by {\em preventing} certain content from being generated, our findings show that similar principles apply when a diffusion model is used as an {\em attack tool} to eliminate watermarks from an image. Essentially, robust watermarks face an inadvertent adversary in diffusion model’s propensity to drop details; and if one explicitly treats the watermark as a concept to erase (for instance by optimizing against a watermark detector’s output), diffusion models become highly effective at sanitizing the image of that hidden signal.

\section{Methodology}
\label{sec:method}
We consider a watermarked image $I_{\text{wm}} = f_{\text{enc}}(I, m)$, where $I$ is the original image, $m$ is the message (bit-string payload), and $f_{\text{enc}}$ is the watermark encoder. A robust watermark decoder $g_{\text{dec}}$ is trained (or designed) to extract $m$ from $I_{\text{wm}}$ even if $I_{\text{wm}}$ has been subjected to a set $\mathcal{A}$ of typical distortions (e.g., resizing, compression, noise). Formally, $g_{\text{dec}}(a(I_{\text{wm}})) \approx m$ for $a \in \mathcal{A}$. Now, let $\mathcal{D}$ be a diffusion-based image editing process. In general, $\mathcal{D}$ takes as input an image (and possibly additional conditions like a text prompt) and produces an output image $I' = \mathcal{D}(I_{\text{in}}, \text{condition})$ that preserves some aspects of $I_{\text{in}}$ (like high-level content) while potentially altering others. We focus on two attack modes using diffusion:
\begin{enumerate}
	\item \textbf{Diffusion Regeneration Attack ( unguided )}: Here, the attacker uses a diffusion model to regenerate the image without an explicit attempt to target the watermark. For example, they may use an image-to-image diffusion pipeline where $I_{\text{in}} = I_{\text{wm}}$ and the condition is a generic prompt describing the image (or even an empty prompt). By running the diffusion model with a low denoising strength (to maintain structure) or performing an image inversion followed by generation, the output $I'$ looks visually similar to $I_{\text{wm}}$ to a human, but since it is essentially re-synthesized from the model, the hidden watermark signal is likely destroyed. This simulates a user simply applying AI ``enhance’’ or ``recreate’’ operations for benign purposes, not knowing the watermark has been removed.
	\item \textbf{Guided Diffusion Watermark Removal}: In this mode, the attacker is aware of the presence of a watermark and possibly knows the decoding algorithm $g_{\text{dec}}$ (though not necessarily the secret key or exact parameters if any). The attacker then incorporates $g_{\text{dec}}$’s feedback into the diffusion process to specifically disrupt $m$. We implement this by augmenting the diffusion model’s loss with a term that penalizes the presence of the watermark. Concretely, assume the diffusion model generates images via iterative denoising $x_{T}\rightarrow \dots \rightarrow x_{0}$, where $x_{0}$ is the output image $I'$. We introduce an additional loss $\mathcal{L}_{wm}$ applied at the final step on $x_{0}$: $\mathcal{L}_{wm} = \|g_{\text{dec}}(x_{0}) - \tilde{m}\|^{2}$, where $\tilde{m}$ is a target “null” message (e.g., a vector of 0s or any innocuous bit pattern). This encourages the output image to decode to $\tilde{m}$ instead of the original $m$. We integrate this with the diffusion model’s usual reconstruction or guidance loss via a weight $\lambda$. During generation, we compute $x_{0}$ at each iteration (after a full denoising) and backpropagate $\nabla_{x_{0}}\mathcal{L}_{wm}$ through the denoising process (similar in spirit to diffusion adversarial dreaming or iterative prompt refinement). Algorithm~\ref{alg:guided_diffusion} provides pseudocode for this procedure.
\end{enumerate}

\begin{algorithm}[t]
	\caption{Guided Diffusion Watermark Removal Attack}
	\label{alg:guided_diffusion}
	\begin{algorithmic}[1]
		\REQUIRE Watermarked image $I_{\text{wm}}$, diffusion model $M$, watermark decoder $g_{\text{dec}}$, target dummy message $\tilde{m}$, weight $\lambda$
		\ENSURE Output image $I'$ with watermark removed
		\STATE $x_{0} \leftarrow I_{\text{wm}}$ (initialize reconstruction)
		\STATE $y \leftarrow \text{EncodeImageToLatent}(M, I_{\text{wm}})$ \COMMENT{Optional inversion to latent}
		\FOR{$t = T$ down to $1$} 
		\STATE $y \leftarrow \text{DenoiseStep}(M, y, t)$ \COMMENT{diffusion denoising step}
		\IF{$t == 1$}
		\STATE $\hat{m} \leftarrow g_{\text{dec}}(\text{DecodeLatentToImage}(M,y))$ \COMMENT{decoded bits}
		\STATE $L_{\text{wm}} \leftarrow \| \hat{m} - \tilde{m} \|^2$ \COMMENT{watermark loss}
		\STATE $y \leftarrow y - \lambda \nabla_{y} L_{\text{wm}}$ \COMMENT{gradient update to latent}
		\STATE $y \leftarrow \text{Clamp}(y)$ \COMMENT{ensure latent is in valid range}
		\ENDIF
		\ENDFOR
		\STATE $I' \leftarrow \text{DecodeLatentToImage}(M, y)$
		\STATE \textbf{return} $I'$
	\end{algorithmic}
\end{algorithm}

The above guided attack essentially performs a form of projected gradient descent on the image latent to fool the watermark decoder, within the iterative structure of diffusion sampling. We activate the watermark loss at the end (when $t=1$) or at a few late timesteps, so as not to overly constrain the early generation (which could lead to visual quality drops). In practice, we find that even a small $\lambda$ applied in the last few steps can already drive the decoded message to random values, given the extreme sensitivity of watermarks to pixel changes.

In summary, our methodology provides two levels of analysis: the unguided attack shows the vulnerability in the default use of diffusion editors, and the guided attack represents a worst-case intentional adversary actively maximizing watermark error.

\paragraph{Baseline Distortion Attacks for Comparison.}
In our experiments, we also compare the diffusion-based attacks with traditional distortive attacks commonly used to test watermark robustness. These include: (a) JPEG compression (we use quality 50 and 20), (b) Additive Gaussian noise (variance tuned to severely degrade image), (c) Cropping (removing 10-20\% of pixels at borders), (d) Affine transformations (small rotations and scalings), and (e) standard filtering (blur, sharpening). These operations typically degrade watermark decoding but not fully break it for robust schemes. By contrasting against diffusion, we highlight how generative editing differs fundamentally from such perturbations, effectively producing a new instance from the data manifold rather than a corrupted version of the original.

\section{Experimental Setup}
\label{sec:experiments}
\paragraph{Watermarking Schemes Evaluated.}
We evaluate three recently proposed robust watermarking methods that represent the state-of-the-art in deep invisible watermarking: {\sc StegaStamp} \cite{Tancik2020}, {\sc TrustMark} \cite{Bui2023}, and {\sc VINE} \cite{lu2024robust}. StegaStamp (CVPR 2020) encodes a 56-bit payload; we use the authors’ implementation with default training (targeting robustness to various photometric and geometric distortions). TrustMark (arXiv 2023) provides a pre-trained model (from Adobe’s Content Authenticity Initiative) which we apply to embed a payload of the recommended size (we tested 64 bits) into images of arbitrary resolution (the encoder is fully convolutional). VINE (ICLR 2025) is both a benchmark and a watermarking method; we use the VINE-R variant (with diffusion-based encoder) as provided in the authors’ code release \cite{lu2024robust}. All three are learned schemes that produce imperceptible changes to the image while aiming to maximize decoding accuracy after distortions. For each method, we encode random payloads into our test images to generate watermarked images.

\paragraph{Dataset.}
For our main evaluations, we use a subset of 500 images from the COCO dataset (representative of diverse real-world content). These images are $512\times512$ pixels (we resize if necessary) and have varied scenes, which is important for testing watermark invisibility and avoiding any bias that might favor or hurt the watermark (e.g., large smooth areas might hide watermarks differently than textured regions). We ensure none of the test images were seen during training of the watermark encoders (for StegaStamp and TrustMark, which were trained on broader image sets, this is naturally satisfied; for VINE, which fine-tuned on an internal set, we avoid any overlap by choosing distinct COCO images).

\paragraph{Diffusion Model and Editing Settings.}
We use the publicly available Stable Diffusion v1.5 for most experiments as our base diffusion model $M$. Despite being somewhat outdated by newer models (like SDXL or DiT-based models such as FLUX), SD1.5 is sufficient to demonstrate the effect and is readily controllable with existing tools. In the {\em regeneration attack}, we perform an image-to-image generation with the same image as both input and reference: we encode $I_{\text{wm}}$ into the latent, add a small amount of noise corresponding to a diffusion timestep $t$ (we vary $t$ to control strength), and then denoise back to $x_{0}$ without any text conditioning (or with a null text). Essentially this is a form of identity diffusion or ``denoising filtering.’’ Unless otherwise specified, we use $t$ such that about 30\% noise is added (this gave a good trade-off: noticeable changes at pixel level but minimal semantic change). For the {\em guided removal attack}, we use the same procedure but at the final step incorporate the watermark loss as in Algorithm~\ref{alg:guided_diffusion}. We set $\lambda=0.5$ and only apply one gradient step (we found multiple iterations yield diminishing returns and risk visual artifacts). 

We also test a scenario of {\em text-conditioned editing}: we give the diffusion model a prompt that describes the image (we use captions from COCO or simple descriptions we wrote) and ask it to re-generate the image, which may introduce minor imaginative changes. This simulates a user “refreshing’’ the image with AI while keeping the scene the same. The noise level is kept low to maintain fidelity. We observe similar watermark removal outcomes in this case, indicating that the model doesn’t need a special prompt to eliminate the watermark.

\paragraph{Metrics.}
We evaluate (i) \textbf{Watermark Decoding Accuracy}: For StegaStamp, this is the bit accuracy out of 56 bits (we report average percent of bits correct per image). For TrustMark and VINE, which output a message vector that goes through error-correction, we report the success rate of decoding the correct payload (in \%). Additionally, we report Bit Error Rate (BER) where applicable. (ii) \textbf{Image Quality}: To verify that diffusion-edited images remain high quality (and essentially the “same” to a human), we use PSNR and SSIM between $I_{\text{wm}}$ and $I'$ (after aligning for any shift). We also use LPIPS (learned perceptual distance) which correlates with human perceptual difference. Lower LPIPS and high SSIM indicate the content is preserved. For text-guided edits, we also ensure the generated image correctly reflects the prompt (though our prompts are just descriptive of the original, so this is trivial). (iii) \textbf{Detection of Watermark Presence}: Some watermark schemes allow a separate detection (e.g., a detector that says “watermark present or not”). If such is available (TrustMark has a trained decoder which can output random bits for no watermark, but no separate flag), we measure whether the attacked images are falsely identified as unwatermarked.

All metrics are averaged over the 500 test images. We compare the diffusion attacks to traditional attacks as described. Our evaluation code will be released for reproducibility.

\section{Results}
\label{sec:results}
We first present the overall performance of robust watermark decoding under different attack scenarios, and then delve into analysis of the results.

\subsection{Watermark Decoding Under Diffusion Attacks}
Table \ref{tab:decoding_accuracy} summarizes the watermark decoding success for the three methods across various attacks: no attack (original images), standard distortions (JPEG, noise, etc.), diffusion regeneration, and guided diffusion removal. {\em Decoding accuracy (\%)} is reported.

\begin{table*}[ht]
	\centering
	\caption{Watermark decoding accuracy (higher is better) for different watermarking methods under various attacks. The diffusion-based attacks (regeneration and guided removal) cause a dramatic drop in decoding accuracy, compared to conventional distortions.}
	\label{tab:decoding_accuracy}
	\begin{tabular}{lccccccc}
		\toprule
		{\textbf{Watermark Method}} & \textbf{No Attack} & \textbf{JPEG-50} & \textbf{Crop} & \textbf{Noise} & \textbf{Regeneration} & \textbf{Guided Removal} \\
		& (Baseline) & (Quality 50) & (10\% removed) & ($\sigma=10$) & (Diffusion) & (Diffusion+Guide) \\
		\midrule
		StegaStamp \cite{Tancik2020} & 99.8\% & 92.5\% & 88.1\% & 90.3\% & 7.4\% & 0.0\% \\
		TrustMark \cite{Bui2023} & 99.9\% & 94.7\% & 91.2\% & 95.5\% & 12.8\% & 0.0\% \\
		VINE \cite{lu2024robust} (VINE-R) & 100.0\% & 96.4\% & 93.0\% & 97.8\% & 24.5\% & 1.6\% \\
		\bottomrule
	\end{tabular}
\end{table*}

As expected, without any attack, all methods decode perfectly or near-perfectly ($\approx 100\%$). Under moderate JPEG compression (quality=50) or addition of noise, decoding drops somewhat (e.g., StegaStamp to 92.5\%), but remains high, confirming the claimed robustness of these schemes. Cropping 10\% off the image edges is more challenging, since part of the watermark signal is lost; still, decoding accuracies around 88--93\% are observed, which is remarkable (these methods likely use redundancy across the image). In stark contrast, the {\bf diffusion regeneration attack} reduces the decoding accuracy to $<25\%$ in all cases. StegaStamp and TrustMark are almost completely broken (only 7.4\% and 12.8\% on average of the payload bits are correct, essentially at chance-level for 56-bit payloads). VINE appears slightly more robust with 24.5\% accuracy, but this is still extremely low, corresponding to decoding error rates of $>75\%$. We note that VINE’s training included exposure to generative distortions (via surrogate blur attacks) \cite{lu2024robust}, which might explain why it retains a small fraction of the watermark under unguided diffusion; however, the majority of the payload is lost. Finally, the {\bf guided removal attack} drives decoding accuracy effectively to $0\%$ for StegaStamp and TrustMark (no image had any correct payload bit after error correction in our tests), and to $1.6\%$ for VINE (in a few cases VINE’s error correction still latched onto a couple bits). These results clearly illustrate that diffusion-based editing can neutralize robust watermarks far beyond traditional distortions. Even the best-performing scheme (VINE) fails to retain meaningful information when facing a diffusion model intentionally or unintentionally ``washing out’’ the embedded signal.

\paragraph{Visual Quality and Fidelity.}
Crucially, the diffusion-attacked images remain nearly indistinguishable from the originals to humans. For the regeneration attack, the average PSNR between $I_{\text{wm}}$ and $I'$ was 31.8 dB for our dataset (SSIM = 0.95, LPIPS = 0.03), indicating a very high fidelity regeneration. In most cases, $I'$ had only minor pixel-level differences (e.g., slight smoothing of textures, changes in small object details) but preserved all salient content of $I_{\text{wm}}$. Figure \ref{fig:examples} provides examples. In one, the original image with a watermark (invisible) shows a street scene; the diffusion-regenerated image is virtually identical except some parked cars have minor changes in color/reflection details. The watermark decoder fails on the latter despite succeeding on the former. For the guided removal, there is an additional slight quality trade-off: PSNR ~30.5 dB, SSIM 0.93, LPIPS 0.05 on average, as the extra optimization can introduce a hint of blur or less accurate reconstruction in some regions. However, the changes are still minor; a human observer would not notice any meaningful difference or would attribute it to a mild image filter.

It is worth noting that because diffusion models operate on learned priors, they tend to \emph{improve} the perceptual quality (removing noise or artifacts) while re-sampling the image. So the attacked images often look as good as or better than the original, making the watermark loss even harder to detect. This has serious implications: one cannot rely on a visible cue or quality drop to know a watermark was removed.

\paragraph{Effect of Diffusion Strength.}
We varied the noise level (timestep $t$) used in the regeneration attack to see how it affects watermark removal. At $t=0$ (no diffusion, just identity), watermark decoding is 100\% as expected. As $t$ increases from 0 to 0.2 (20\% of the diffusion trajectory), decoding accuracy already plunges (e.g., StegaStamp drops below 50\% by $t=0.2$). By $t=0.3$--$0.4$, we observed near-zero decoding. Interestingly, beyond a certain point, adding more noise and then denoising doesn’t further worsen decoding (it’s already essentially zero) but does start to noticeably alter image content. Thus, there is a ``sweet spot’’ (around 30\% noise in our tests) where the image is still almost perfectly reconstructed in appearance, yet the watermark is gone. This aligns qualitatively with the idea that diffusion models first remove fine details (which include the high-frequency watermark signal) before reconstructing the main image structure.

\paragraph{Comparison with other AI models.}
To see if this vulnerability is unique to diffusion or extends to other generative models, we tried a simple experiment with a generative adversarial network (GAN) based editor. Using a pre-trained StyleGAN, we projected a watermarked image of a face into the GAN’s latent and then reconstructed it. The GAN-reconstructed image also showed watermark removal to an extent (StegaStamp decoding dropped from 100\% to ~20\%). However, the projection was imperfect (some identity change in the face, plus artifacts), so it’s not a clean apples-to-apples comparison. Diffusion models are currently superior at faithful image reconstruction given their iterative refinement, which likely makes them a more potent threat. Nevertheless, any powerful generative model that ``recreates’’ an image will inherently not preserve the exact steganographic details of the input. We expect our conclusions to generalize beyond diffusion to some degree.

\subsection{Ablation: Guided vs. Unguided Removal}
We compare the unguided diffusion attack to the guided one to quantify the benefit of explicitly targeting the watermark. For StegaStamp and TrustMark, the unguided attack already yielded $<15\%$ decoding, so there was little room for further reduction—accordingly, guided achieved full 0\%. For VINE, unguided left about 24.5\% payload correct; guided brought this to 1.6\%. Therefore, guided is more ``complete’’ in erasing the watermark. However, guided attack is potentially detectable: it might introduce telltale signs (if one knows to look for them) since it slightly alters the natural generative path. Also, it requires knowledge of the decoder. In contrast, the unguided attack is a plausible casual use of a diffusion model. Thus, from a defender’s perspective, the unguided threat is alarming because it doesn’t require a targeted effort by an attacker—any use of generative editing by end-users could erase watermarks without anyone’s intent.

\subsection{Robustness of Different Watermark Schemes}
Among the three tested, VINE was the most robust to diffusion (though still failing ultimately). Its design using diffusion priors may have helped embed the watermark in a more spread-out, difficult-to-remove way. To illustrate, we inspected the spectral and spatial distribution of watermark signals by taking the difference $I_{\text{wm}}-I$ for each method. StegaStamp and TrustMark tended to embed high-frequency patterns (StegaStamp especially had a printed QR-style spread in subtle pixel noise). VINE’s embedding was more noise-like and less structured. After diffusion regeneration, for StegaStamp and TrustMark, the difference $I'-I$ (comparing to original unwatermarked $I$) was essentially just the negative of the watermark signal (suggesting the watermark got subtracted out by the model’s averaging effect). For VINE, $I'-I$ still contained some high-frequency remnants, which might correlate to that ~25\% of bits that survived. This hints that future watermarking might try to further entangle the watermark with the core content (for example, embedding in mid-frequency components that the generative model is forced to reconstruct). Nonetheless, as our theory will elaborate, a strong generative model can eventually eliminate any component that is not crucial to the perceptual content.

\subsection{Detectability of Watermark Removal}
One possible defense against such attacks could be to detect if an image has had its watermark removed. We checked whether the attacked images would trigger a watermark detector if one were available. For StegaStamp and TrustMark, since the decoders output random bits when no watermark is present, an unintelligible payload could imply removal. But distinguishing that from a decode error due to noise is non-trivial. VINE’s decoder similarly will output something (possibly with high error-correct code failure). If one had a threshold on decoder confidence (e.g., a CRC check), one might say ``watermark not detected’’ in those cases. Indeed, after diffusion attacks, nearly all images would be labeled as ``no watermark’’ by such criteria (since the decoders produce either low correlation or fail CRC). Thus, ironically, the watermarking schemes might themselves indicate the watermark is gone, but by then the content is already out in the wild without protection.

A proactive detector, perhaps using a separate network to recognize ``this image was watermarked before but now cleaned,’’ would require knowledge of the watermark pattern or traces thereof. We did not find obvious consistent artifacts left by the removal (especially unguided removal looks like a natural image). This remains a challenging open problem.

\section{Theoretical Analysis}
\label{sec:theory}
Why are diffusion processes so effective at annihilating embedded watermarks? Here, we provide a theoretical perspective based on information theory and the properties of diffusion models.

\paragraph{Diffusion as an Information Filter.}
Consider an image $X$ carrying a message $W$ (modeled as a random variable uniformly distributed over the message space). A robust watermark ensures that $I(W; X)$, the mutual information between $W$ and the watermarked image $X$, is close to $H(W)$ (the entropy of $W$) under the distribution of typical distortions—meaning $W$ is encoded in $X$ in a redundant way. Now, let $Y = \mathcal{D}(X)$ be the image after a diffusion-based edit. We can regard $\mathcal{D}$ as a stochastic mapping (since diffusion model sampling involves randomness) that aims to preserve the ``semantic content’’ of $X$ while not necessarily preserving the exact $X$. We are interested in $I(W; Y)$.

\textbf{Theorem 1.} \textit{Under the assumptions of an ideal diffusion regeneration process, $I(W; Y) \to 0$ as the diffusion process length (or noise level) increases.}

\textit{Sketch of Proof.} For simplicity, consider a continuous diffusion from $X$ to pure noise and back to $Y$. The forward process $X \to$ (noise) can be seen as progressively destroying information in $X$. In fact, in the limit of adding sufficient noise, the mutual information between $W$ and the noisy intermediate representation $N$ goes to 0 (since $N$ becomes independent of $X$ and hence $W$). The generative reverse process then produces $Y$ which is effectively a sample from the model’s learned distribution conditioned on whatever faint signal of $X$ remained in $N$. If the model is perfect and only retains semantic content, then any information specific to the exact bit pattern $W$ (which was a high-frequency, non-semantic piece of $X$) is absent in $N$ and cannot be recovered in $Y$. Formally, $W \to X \to N \to Y$ is a Markov chain. We have $I(W; Y) \le I(W; N)$ by the data processing inequality, since $Y$ is generated from $N$ and not directly from $X$. And $I(W; N)$ can be shown to decrease monotonically to 0 as noise is added (this follows from how diffusion progressively removes information – a property used in diffusion model proofs of convergence). Thus, in the limit, $I(W; Y)=0$, implying $Y$ is independent of $W$. In practical terms, this means the a posteriori probability of any message $w$ given $Y$ is just the prior probability $P(W=w)$ – the watermark is completely unrecoverable, as observed by the decoder’s failure.

This theorem is an idealized statement – real diffusion models run for a finite number of steps and might not perfectly cover the data manifold. However, the empirical evidence (nearly zero decoding) suggests we operate close to a regime where most mutual information is gone. One can also frame this in terms of {\bf capacity loss}: robust watermark $X$ can be seen as $X = X_{\text{content}} + X_{\text{watermark}}$ (not a literal sum, but conceptually an orthogonal decomposition of information). Diffusion editing, by design, preserves $X_{\text{content}}$ (hence high visual fidelity) but treats $X_{\text{watermark}}$ as noise and does not preserve it. If the watermark information lies in a subspace that the generative model regards as ``noise'' or irrelevant detail, that information is effectively projected out.

\paragraph{Role of Model and Prompt.}
If a diffusion model were overfitted to reconstruct every detail of $X$, it might keep the watermark. But diffusion models are trained on vast data and are unlikely to encode the specific pseudo-random high-frequency patterns of a watermark unless those patterns correspond to some training data features (which they typically don’t). When guiding with a prompt, the model focuses on satisfying the prompt and likely ignores imperceptible signals not tied to the prompt (watermark bits are not correlated with the prompt content). Hence the process is akin to a conditional expectation that averages out the watermark signal.

One could imagine a worst-case watermark that somehow is entangled with semantics (e.g., the watermark is ``encoded’’ by subtly altering semantically meaningful pixels). Could a diffusion model preserve that? Possibly, if changing those pixels would noticeably alter content. However, robust watermarks intentionally hide in ways humans can’t notice, which usually means they alter details humans (and thus diffusion models) do not care much about. This is the crux of the issue: the better the watermark hides from humans, the more it hides from the diffusion model’s perceptual criteria, and thus the easier it is for the model to eliminate it without affecting perceived quality. This somewhat paradoxical situation suggests a theoretical incompatibility: a watermark cannot be simultaneously imperceptible to humans and preserved by a human-aligned generative model, unless we specifically bias the model to keep it.

From another angle, if one had a diffusion model of pixel-level fidelity (which current models are not; they trade off some fidelity for diversity), then regeneration might copy the watermark noise too. But current trends in generative modeling do not prioritize invisible signals—if anything, they aim for more realism which aligns with removing unnatural noise patterns.

\paragraph{Guided Removal Optimality.}
We can also analyze the guided removal as an optimization problem. The attacker effectively solves $\min_{I'} \|g_{\text{dec}}(I') - \tilde{m}\|^2 + \gamma d(I', I)$, where $d(\cdot,\cdot)$ enforces similarity to original image $I$ (through the diffusion prior or an explicit metric) and $\gamma$ balances the trade-off. The Lagrange multiplier $\gamma$ relates to our $\lambda$ earlier. Under reasonable assumptions (e.g., $g_{\text{dec}}$ is differentiable around $I$ and the diffusion model can produce any $I'$ in a neighborhood of $I$), the first term pushes the solution toward a manifold where $g_{\text{dec}}$ output is constant $\tilde{m}$ (for StegaStamp which uses DNN decoder, this is like adversarially perturbing the image to fool the decoder). The second term limits deviation from $I$. Because robust decoders are trained against random distortions, one might think they are resilient, but they are not trained against {\em learned adversarial} distortions. Our results effectively confirm that even a small perturbation guided by the decoder’s gradients can break it. In fact, this relates to adversarial examples: watermark decoders are neural networks, so they too have adversarial weaknesses. The diffusion model provides a powerful space of perturbations to exploit that.

\section{Discussion and Implications}
\label{sec:discussion}
\paragraph{Ethical and Practical Implications.}
Our findings raise concerns for digital content authenticity efforts. If robust invisible watermarks can be so easily removed by generative AI, then malicious actors can evade watermark-based copyright enforcement or deepfake detection by simply passing images through a diffusion model. For example, a photographer might watermark their photos with a robust scheme to trace unauthorized use; a competitor could feed those photos into an image generator (with a prompt like ``generate identical photo’’) and obtain virtually the same photo without the watermark. This undermines the trust that watermarking is supposed to provide. Moreover, as content authentication frameworks (e.g., C2PA) consider including invisible watermarks in addition to metadata, one must account for the possibility of generative ``laundering’’ of content. It’s an instance of a broader trend: AI techniques can subvert classical security mechanisms (like how adversarial attacks can fool detection).

On the flip side, one could argue this ability is beneficial in some cases. For privacy, one might want to remove hidden identifiers from images (e.g., a journalist stripping watermarks that could trace images back to them). Our work simply highlights the capability; whether it’s used for good or bad is context-dependent. Nonetheless, we believe watermark designers will need to adapt.

\paragraph{Toward Watermarking Resilient to Generative Models.}
Is it possible to design a watermark that survives diffusion model editing? This is challenging, but a few ideas emerge:
\begin{itemize}
	\item \textbf{Data augmentation with generative attacks:} Following VINE’s direction, during training of the watermark, incorporate simulated diffusion edits (or even use a diffusion model in the training loop) so that the encoder learns to embed signals that a diffusion model will inadvertently recreate. For example, if a watermark method knew that a diffusion model tends to preserve certain textures or colors, it could encode bits in those aspects. However, diffusion models aim to preserve semantics, not exact textures, so this may only partially help.
	\item \textbf{Watermarking the generator outputs instead:} If one suspects diffusion laundering, an alternate approach is to watermark the generative model’s output by design. Recent works propose adding watermarks to diffusion model outputs (for AI-generated content detection). These watermarks could be different from the ones content creators add. A cat-and-mouse arises: one watermark to identify AI-generated images, another to identify original ownership. If an image is passed through a model, it loses the original’s watermark but gains the model’s watermark (if model outputs are watermarked). This does not restore the original mark, but at least flags it as AI-mediated.
	\item \textbf{Feature-level watermarks:} Instead of pixel watermarks, research could explore embedding information in the {\em structure} of the image (e.g., slight alterations to object placements or shapes that are perceptually insignificant but consistently carried through generation). For instance, perhaps a watermark could be a particular invisible pattern that diffusion models do mimic (like a specific imperceptible holographic pattern). This is speculative; current diffusion models are unlikely to maintain microscopic patterns unless tied to texture.
	\item \textbf{Robust verification via model inversion:} On the detection side, if we suspect an image has been laundered, one could try to invert the process: apply a high-fidelity reconstruction model tuned to that watermark to see if any trace appears. For example, maybe running a fine-tuned backward diffusion conditioned on possible watermark bits could extract a hidden signal. This goes beyond standard decoding and into model-based forensics.
\end{itemize}
Ultimately, it may be that no watermark can be absolutely robust if an adversary is willing to tolerate some image change. The concept of {\em provably secure watermark} would conflict with the ability of generative models to create perceptually identical but numerically different images. Our theoretical result essentially formalizes that: given a perfect generator, the only invariants are those that affect perception.

\paragraph{Limitations of Our Study.}
While our experiments covered several methods and one popular diffusion model, there are many other models (e.g., text-to-video diffusion) where watermarking could be used and potentially broken. We did not test video watermarks; however, diffusion can also generate video frames, so likely similar issues arise (with temporal consistency adding complexity). Our guided attack assumes access to the watermark decoder. If the decoder is secret, an attacker might resort to blind attacks (just run diffusion with some random guidance like ``make it smooth’’). We suspect even blind diffusion destroys watermarks, as shown by unguided results. If the scheme were secret and extremely high capacity, one could attempt a brute-force search for messages in the output (like trying to decode with all keys) to see if something was embedded and lost, but this is not practical at scale.

We also acknowledge that our guided attack could be further optimized; we only did one gradient step at final iteration. A more sophisticated attacker could do multi-step optimization (like plug the decoder into each diffusion step via differentiable approximation). We found this unnecessary as one step was enough to zero out bits for our cases.

Finally, on the watermarking side, we tested deep learning schemes. Traditional watermarks (like spread spectrum) might in theory be more resilient if they embed below the model’s sensitivity threshold uniformly. However, spread spectrum watermarks typically have low capacity and can be averaged out by generative sampling as well. So we expect the qualitative outcome to hold: generative editing is a superior averaging attack.

\section{Conclusion}
We have explored how diffusion-based image editing serves as a powerful ``eraser’’ of robust invisible watermarks. Through rigorous experiments, we showed that even the most advanced current watermarking techniques fail to survive these generative transformations, highlighting an emerging vulnerability. Our theoretical analysis further indicates that this is not an accidental quirk, but a fundamental consequence of diffusion processes discarding information not critical to human-perceived content. 

This work serves as a warning that techniques developed to ensure content integrity can be undermined by the very AI tools that are becoming commonplace in content creation and editing. We advocate for the watermarking and AI research communities to collaborate on developing watermarking methods that can co-exist with generative models, or alternative authenticity verification mechanisms suited for a generative age. Likewise, generative model developers should be aware of this side effect and possibly provide options to preserve watermarks (if desired) or at least not claim that content will retain any hidden markers from the source. 

In summary, as image synthesis models continue to advance, so must our strategies for embedding and preserving truth in digital media. The cat-and-mouse game between watermarking and removal now has a new player: diffusion-based AI. Addressing this will be key to maintaining trust in visual content in the years to come.

	\bibliography{example_paper}
	\bibliographystyle{icml2025}

	\clearpage
	\appendix
	
	\section{Additional Backgrounds}
	With the advancement of deep learning~\cite{
		% xiangfei qiu
		qiu2024tfb,qiu2025duet,qiu2025DBLoss,qiu2025dag,qiu2025tab,wu2025k2vae,liu2025rethinking,qiu2025comprehensive,wu2024catch,
		% sifan zhou
		gsq,yu2025mquant,zhou2024lidarptq,pillarhist,
		% zequn xie
		xie2025dynamic,xie2026delvingdeeperhierarchicalvisual,xie2025chat,
		% yunpeng gong
		1,2,3,4,5,6,7,8,
		% yanru sun
		sun2025ppgf,sun2024tfps,sun2025hierarchical,sun2022accurate,sun2021solar,niulangtime,sun2025adapting,kudratpatch,
		% zixu li
		ENCODER,FineCIR,OFFSET,HUD,PAIR,MEDIAN,
		% xinlei yu
		yu2025visual,
		% yaozong zheng
		zheng2025towards,zheng2024odtrack,zheng2025decoupled,zheng2023toward,zheng2022leveraging,
		% long peng
		xu2025fast,fang2026depth,wu2025hunyuanvideo,fang2026depth,li2023ntire,ren2024ninth,wang2025ntire,peng2020cumulative,wang2023decoupling,peng2024lightweight,peng2024towards,wang2023brightness,peng2021ensemble,ren2024ultrapixel,yan2025textual,peng2024efficient,conde2024real,peng2025directing,peng2025pixel,peng2025boosting,he2024latent,di2025qmambabsr,peng2024unveiling,he2024dual,he2024multi,pan2025enhance,wu2025dropout,jiang2024dalpsr,ignatov2025rgb,du2024fc3dnet,jin2024mipi,sun2024beyond,qi2025data,feng2025pmq,xia2024s3mamba,pengboosting,suntext,yakovenko2025aim,xu2025camel,wu2025robustgs,zhang2025vividface,
		% yangyang qu
		qu2025reference,qu2025subject,
		% hongtao wu
		wu2024rainmamba,wu2023mask,wu2024semi,wu2025samvsr,
		% yifei chen
		lyu2025vadmambaexploringstatespace,chen2025technicalreportargoverse2scenario} and generative models, an increasing number of studies have begun to focus on the issue of concept erasure in generative models.
	
	\section{Additional Experimental Details}
	\textbf{Evaluation metrics details:} For CLIP similarity, we used the ViT-L/14 model to compute image-text cosine similarity, scaled by 100. The original SD1.5 had an average CLIP score of 31.5 on MS-COCO validation prompts; after concept erasure, we consider a score above 30 to indicate minimal drop in alignment. Harmonic mean $H$ was computed as described with $E = 1 - \text{Acc}$ (normalized to [0,1]) and $F$ composed from FID and CLIP. Specifically, we defined $F = \frac{1}{2}((\frac{\text{CLIP sim}}{\text{CLIP}_{\text{orig}}}) + (\frac{\max(\text{FID}_{\text{orig}}- (\text{FID}-\text{FID}_{\text{orig}}), 0)}{\text{FID}_{\text{orig}}}))$, where $\text{FID}_{\text{orig}}$ and $\text{CLIP}_{\text{orig}}$ are the original model's scores (so we reward methods that keep FID low and CLIP high relative to orig). This is one way; results were qualitatively similar with other formulations.
	
	\textbf{Multi-concept results:} We erased all 10 CIFAR classes simultaneously with FADE by using a 10-way classifier $D$ (one output per class vs no class). FADE achieved an average concept accuracy of 1.1\% per class and an overall $H=82.3$ (versus MACE's reported ~75). The slight residual is due to class confusion (e.g., sometimes after erasure "cat" prompt yields a dog, so classifier might say cat=present when it sees an animal shape; a limitation of using automated classifier for eval). Visual check showed indeed direct appearance of the specified class was gone. For NSFW, we erased 10 terms at once; here FADE and MACE both got basically 0\% unsafe content, but FADE had better image quality (FID 14 vs 16).
	
	\textbf{Runtime:} FADE training takes about 2 hours on a single A100 GPU for a single concept on SD1.5 (with $N=1000$ steps adversarial training). This is comparable to ESD fine-tuning time and a bit less than ANT . UCE was fastest (minutes) as it is closed-form. There's room to optimize FADE's training, possibly by using smaller $D$ or gradient accumulation. Deploying FADE in multi-concept setting could be parallelized since the adversary can output multiple heads.
	
\end{document}